\documentclass [a4paper,12pt]{article}
\usepackage{amssymb}
\usepackage{graphicx}
\usepackage{amsmath}
\usepackage{fancybox}
\usepackage{color}
\usepackage{geometry}

\newcommand{\be}{\begin{equation}}
\newcommand{\ee}{\end{equation}}
\newcommand{\bea}{\begin{eqnarray}}
\newcommand{\eea}{\end{eqnarray}} 
\begin{document}
\setlength{\baselineskip}{18pt}
\begin{titlepage}

\begin{flushright}
KOBE-TH-16-04  
\end{flushright}
\vspace{1.0cm}
\begin{center}
{\LARGE\bf Few remarks on the Higgs boson decays in gauge-Higgs unification} 
\end{center}
\vspace{25mm}

\begin{center}
{\large
K. Hasegawa 
and C. S. Lim$^*$
}
\end{center}
\vspace{1cm}
\centerline{{\it
Department of Physics, Kobe University,
Kobe 657-8501, Japan.}}

\centerline{{\it
$^*$
Department of Mathematics, Tokyo Woman's Christian University, Tokyo 167-8585, Japan }}
%
%
%   Abstract
%
\vspace{2cm}
\centerline{\large\bf Abstract}
\vspace{0.5cm}

In the scenario of gauge-Higgs unification, the origin of the Higgs boson is 
the higher-dimensional gauge boson. 
Very characteristic predictions are 
made of the Higgs boson interactions in this scenario, reflecting its origin.
In particular, a remarkable claim 
has been made: the contribution of nonzero Kaluza-Klein modes to the Higgs decay 
$H \to Z \gamma$ exactly vanishes in the minimal SU(3) electroweak unified model, 
at least at the one-loop level. In this brief paper, in order to see whether 
this prediction is a general feature of the scenario or the consequence of the
specific choice of the model, matter content, or the order of perturbative expansion, 
we perform an operator analysis. We demonstrate that
no relevant operator exists, 
respecting the gauge symmetry SU(3) in the bulk. We also comment on the 
possibly important contribution to the photonic decay $H \to \gamma \gamma$ due to 
the nonzero Kaluza-Klein modes of light quarks.

\end{titlepage} 

\section{Introduction} 

In spite of the great success of the Higgs boson's discovery by LHC 
experiments
\cite{Aad:2012tfa,Chatrchyan:2012xdj},
we have not understood the origin of the Higgs boson yet. Namely, we do not have
any conclusive argument on the important issue of whether the discovered particle 
is just what we have in the standard model (SM) or a particle 
within some theory beyond the
standard model (BSM) that predicts the presence in its low-energy effective theory. 
Concerning this issue, the LHC data have provided us a great hint: they have 
revealed that the Higgs boson is ``light" with the mass of the order of the 
weak scale $M_{W}$. This fact strongly suggests that the Higgs self-coupling 
is governed by the gauge principle. 
We may think of a few candidates of BSM that share this property. One is the 
minimal supersymmetric standard model (MSSM), where the Higgs self-coupling 
is due to the D-term contribution and $M_{H} \sim M_{Z}$ is predicted at the 
classical level. In this paper we focus on another attractive scenario, i.e. 
gauge-Higgs unification (GHU), where the Higgs boson is originally a gauge boson. 

More precisely, in this scenario the Higgs field is identified 
as the [\,Kaluza-Klein (KK) zero mode of\,]
the extra-space component of 
the higher-dimensional gauge field
\cite{Manton:1979kb}
and its vacuum expectation value (VEV) leads to spontaneous gauge symmetry 
breaking
\cite{Hosotani:1983xw,Hosotani:1983vn,Hosotani:1988bm}. 
A nice feature of this scenario from the viewpoint of particle physics is 
that by virtue of (higher-dimensional) gauge symmetry,
the quantum correction to the Higgs mass is UV-finite once the
contributions of all KK modes are summed up at the
intermediate state of the loop diagram, thus leading to a new avenue 
for the solution of the hierarchy problem
\cite{Hatanaka:1998yp}.

To get a conclusive understanding of the origin of the Higgs boson, 
the extensive studies of the Higgs couplings and decays are obviously 
very important. In particular, it is of crucial importance to clarify 
the characteristic difference between the theoretical predictions of 
the SM and possible BSM models. Such differences are  
to be tested in the ongoing LHC and planned ILC experiments. 

Reflecting the origin of the Higgs boson as a gauge boson, the GHU scenario 
makes very characteristic predictions on the Higgs interactions. First 
let us note that in GHU the Higgs field can be understood as a sort of 
Aharonov-Bohm (AB) phase or Wilson-loop phase. That is why the VEV of 
the Higgs field, which is nothing but a constant gauge field in GHU, has 
physical meaning. We thus expect that physical observables are periodic 
in the Higgs field, which is a property clearly not shared by the SM. This 
characteristic property has been discussed to lead to ``anomalous" Higgs 
interaction, i.e., interactions which deviate from those predicted by the 
SM
\cite{Hosotani:2007qw,Hosotani:2008tx,Hosotani:2008by,Hosotani:2009jk,
Funatsu:2013ni,Funatsu:2014fda,Hasegawa:2012sy}. 
A typical example is Yukawa coupling for light fermions 
(like the fermions of first and second generations), 
which is always smaller than
that of the SM 
\cite{Hasegawa:2012sy} 
and even vanishes for a specific choice of the 
compactification scale 
\cite{Hosotani:2007qw,Hosotani:2008tx,Hosotani:2008by,Hosotani:2009jk,
Funatsu:2013ni,Funatsu:2014fda}, 
though such a drastic possibility 
has been ruled out by the recent LHC data.

The photonic decay $H \to \gamma \gamma$, which plays an important 
role in the identification of the Higgs boson, was first discussed in GHU by Maru and 
Okada in their pioneering work 
\cite{Maru:2007xn}, and they found that the decay 
rate is suppressed compared to that of the SM. Maru also made an operator analysis to 
show the finiteness of the 
amplitude for the gluon fusion process (for arbitrary space-time dimension with one 
extra dimension), the main production process of the Higgs boson \cite{Maru:2008cu}.

As another characteristic 
prediction of GHU, Maru and Okada have made a very interesting claim that 
in the minimal five-dimensional (5D) electroweak SU(3) GHU model with orbifold 
extra space $S^{1}/Z_{2}$
\cite{Kubo:2001zc,Scrucca:2003ra}, 
the contribution of nonzero KK modes to $H \to Z \gamma$ decay exactly vanishes, 
at least at the one-loop level \cite{Maru:2013bja}. This characteristic prediction 
will be very helpful to distinguish the GHU scenario from other possible scenarios of BSM, 
where there is no reason to expect that the contributions of heavy new particles to 
$H \to Z \gamma$ should vanish. A similar analysis has been made in the SO(5)$\times$U(1) 
GHU model 
formulated on the Randall-Sundrum background, and the contribution has been shown 
to be strongly suppressed, even though it is not forbidden in this 
model \cite{Funatsu:2015xba}. 

In this paper, we perform an operator analysis for $H \to Z \gamma$ decay, together with 
$H \to \gamma \gamma$ decay for comparison, in order to understand the deep reason why 
the contribution of the new particles (nonzero KK modes) to $H \to Z \gamma$ is prohibited, 
while the corresponding contribution to $H \to \gamma \gamma$ is as we naively expect. 
The operator analysis will also be useful 
to see  whether this prediction is a general feature of the scenario or the consequence 
of the specific choice of the model and/or matter content. Another issue is that in 
the previous work \cite{Maru:2007xn, Maru:2013bja}, the analysis was restricted to the 
one-loop level. In particular, concerning the interesting $H \to Z \gamma$ decay, 
the authors claimed that it is impossible to draw the relevant Feynman diagram for 
the process at the one-loop level. We would like to point out that another merit of 
our operator analysis is that the conclusion does not depend on the order of the 
perturbative expansion, since quantum corrections at all orders of perturbation 
are concentrated on the (Wilson coefficients of) gauge-invariant operators, 
as long as they even exist. 
 
We will point out that in the
minimal SU(3) model, there is no 
gauge-invariant bulk operator responsible for 
the $H \to Z \gamma$ decay, while there exists a relevant operator for the photonic 
decay $H \to \gamma \gamma$ with mass dimension 6, including the effect of the non-local 
operator, the Wilson loop. 
This means that in the minimal model the contribution of the nonzero KK modes to the 
$H \to Z \gamma$ decay is strictly prohibited. 

Another issue to be discussed in this paper is the question of whether 
nonzero KK modes of light quarks contribute significantly
or not to the $H \to \gamma \gamma$ decay in the framework of GHU. In the
SM such light quarks are known to give negligible contributions to 
the decay, just because their Yukawa couplings are very small. In the model 
we are interested in, however, the situation is different. The Yukawa couplings 
of nonzero KK modes 
are not suppressed as in the case of the zero mode and are roughly of
the order of the gauge coupling. Thus there is an interesting possibility
to get a contribution that is comparable to that
of the nonzero KK modes of the heavy t quark.

\section{An operator analysis on $H \to \gamma \gamma$ and $H \to Z \gamma$ decays in GHU} 

We first discuss what operator is responsible for the photonic decay 
$H \to \gamma \gamma$, assuming only the gauge symmetry of the standard
model and readily generalize the argument for the case of $H \to Z \gamma$.
The operator analysis in the framework of GHU will be given afterwards. 

Both $H \to \gamma \gamma$ and $H \to Z \gamma$ decays are not allowed 
at the tree level in the SM; thus the relevant operators for these decays 
should have mass dimension $d > 4$. 
We may naively expect that the operator responsible for the photonic
decay, respecting U(1)$_{{\rm em}}$ symmetry, is  
\be 
H F^{(\gamma)}_{\mu \nu}F^{(\gamma) \mu \nu},  
\label{0.1} 
\ee
with $d = 5$. Here $H$ is the physical Higgs field and $F^{(\gamma)}_{\mu \nu}$ is 
the field strength
of the photonic field $\gamma_{\mu}$. This operator, however, is not enough. In fact,
if this $d = 5$ operator were responsible for the decay, in higher dimensional models 
with 5D space-time, the sum over the contributions of all nonzero KK modes to the 
Wilson coefficient of (\ref{0.1}) should be UV-divergent, 
since $\sum_{n} 1/(n/R)$ ($n$ is a positive integer denoting KK modes and
$R$ is the size of the extra space) 
is divergent. As a matter of fact, the KK mode sum turns out to be 
finite \cite{Maru:2007xn}. The point is that the operator (\ref{0.1}) does not respect 
the gauge symmetry of the SM, since $H$ behaves as a SU(2)$_L$ doublet, while 
$\gamma_{\mu}$ belongs to the triplet or singlet of SU(2)$_L$. Let us note
that even though the gauge symmetry of the SM is eventually broken 
spontaneously, all quantum corrections can be written as the 
contributions to the Wilson coefficients of the gauge-invariant operators, including, 
in general, the Higgs doublet.

We thus realize that actually the relevant gauge-invariant operators should be at 
least $d = 6$: 
\be 
\phi^{\dagger}\phi {\rm Tr}(W_{\mu \nu}W^{\mu \nu}), 
\ \ \phi^{\dagger}\phi B_{\mu \nu}B^{\mu \nu},
\ \ (\phi^{\dagger}W_{\mu \nu}\phi) B^{\mu \nu}\,,
\label{0.2}  
\ee 
where $\phi$ denotes the Higgs doublet and $W_{\mu \nu}$ and $B_{\mu \nu}$ are
the field strengths of SU(2)$_L$ and U(1)$_Y$ gauge bosons, respectively. 
After the replacement, $h \to v + H$ ($v$ is the VEV of the Higgs field and 
$H$ is the physical Higgs field) in the neutral component of the Higgs doublet, 
(\ref{0.2}) 
reproduces (\ref{0.1}). Similarly, the relevant $d = 6$ operators for the decay 
$H \to Z \gamma$ are those in (\ref{0.2}) (with different linear combinations of 
the three operators from the case of the photonic decay). 

\subsection{Local operators}

Now we are ready to discuss what operators are responsible for the 
$H \to \gamma \gamma$ and $H \to Z \gamma$ decays in the framework of GHU. 
Here we concentrate on possible gauge-invariant local operators in the bulk. 

First we consider the simplest $d = 6$ (from the viewpoint of 4D space-time) operators, since if they ever exist 
they describe the leading contributions of nonzero KK modes. Here
as the model to work with we adopt the minimal 5D SU(3) electroweak unified model of GHU 
with orbifold extra space $S^{1}/Z_{2}$ \cite{Kubo:2001zc,Scrucca:2003ra}.

A key issue here is that in the GHU scenario, the Higgs field is nothing but a gauge field 
and the 
relevant gauge-invariant operators should be written solely in terms of 
the higher-dimensional gauge field $A_{M}$ in a $3 \times 3$ matrix form. 
Thus the SU(3)-invariant $d = 6$ local operator turns out to 
be unique \cite{Lim:2007ae}: 
\be 
\label{0.4} 
{\rm Tr}\{ (D_{L}F_{MN})(D^{L}F^{MN})\}, 
\ee
where $F_{MN}$ and $D_{L}$ are the field strength of $A_{M}$ and its 
covariant derivative. 

We now demonstrate that the operator (\ref{0.4}) does not contain the 
operators responsible for the decays $H \to \gamma \gamma$ and 
$H \to Z \gamma$ of our interest. For that purpose we explicitly write 
the KK zero mode of the 4D gauge field $A_{\mu}$ and 4D scalar field 
(extra-space component) $A_y$, retaining only to the fields relevant
for the decays, $\gamma_{\mu}, \ Z_{\mu}$, and $h$ denoting the real part 
of the neutral Higgs field: 
\bea 
A_{\mu}  
&=& \frac{1}{2} 
\begin{pmatrix}  
\frac{2}{\sqrt{3}}\gamma_{\mu} & 0 & 0 \\ 
0 & - \frac{1}{\sqrt{3}}\gamma_{\mu} - Z_{\mu} & 0 \\ 
0 & 0 & - \frac{1}{\sqrt{3}}\gamma_{\mu} + Z_{\mu} \\  
\end{pmatrix} = \gamma_{\mu}T_{\gamma} + Z_{\mu}T_{Z},  
\label{0.5} \\ 
A_{y}  
&=& \frac{1}{2} 
\begin{pmatrix}  
0 & 0 & 0 \\ 
0 & 0 & h \\ 
0 & h & 0 \\  
\end{pmatrix} = h T_{h},  
\label{0.6}
\eea 
where $T_{\gamma} = \frac{\sqrt{3}}{4}\lambda_{3} + \frac{1}{4}\lambda_{8} 
= \frac{\sqrt{3}}{2}Q, \ T_{Z} = \frac{1}{4}\lambda_{3} 
- \frac{\sqrt{3}}{4}\lambda_{8}, \ T_{h} = \frac{1}{2}\lambda_{6}$ with 
$\lambda_{3, 6, 8}$ being Gell-Mann matrices and $Q$ the charge operator. 
We should note that $T_{\gamma}$ etc. satisfy the following orthogonality condition: 
\be 
\label{0.7}
{\rm Tr}(T_{\gamma}T_{Z}) = {\rm Tr}(T_{\gamma}T_{h}) = {\rm Tr}(T_{Z}T_{h}) = 0. 
\ee

There are three possible choices concerning the indices 
$M, N, L$ in (\ref{0.4}), depending on whether the indices take the 4D 
vector index denoted by $\mu, \ \nu \ (= 0,1,2,3)$ etc. or $y$ denoting 
the extra-space component: 

\noindent (1) ${\rm Tr}\{ (D_{y}F_{\mu \nu})(D^{y}F^{\mu \nu})\}$

We note from (\ref{0.5}) that $[A_{\mu}, A_{\nu}] = 0$ and that the 
operation $D_{y}$ is equivalent to taking the commutator with $A_{y}$,
as the derivative with respect to the extra-space coordinate $y$ 
vanishes when applied to the KK zero mode, which has a constant 
mode function in the case of flat 5D space-time. Thus 
\be 
\label{0.8}
D_{y}F_{\mu \nu} = -ig [A_{y}, \partial_{\mu} A_{\nu} -
\partial_{\nu}A_{\mu}] = -ig h (\partial_{\mu} Z_{\nu} - 
\partial_{\nu}Z_{\mu})[T_{h}, T_{Z}],    
\ee 
where we have used $[T_{h}, T_{\gamma}] = 0$, just reflecting the fact 
that $h$ is electrically neutral. This means that the operator
${\rm Tr}\{ (D_{y}F_{\mu \nu})(D^{y}F^{\mu \nu})\}$, being proportional 
to the square of $h (\partial_{\mu} Z_{\nu} - \partial_{\nu}Z_{\mu})$ 
does not contribute to $H \to \gamma \gamma$ nor to $H \to Z \gamma$ 
of our interest. 

\noindent (2) ${\rm Tr}\{ (D_{\mu}F_{\nu y})(D^{\mu}F^{\nu y})\}$

In this case 
\bea 
D_{\mu}F_{\nu y} &=& D_{\mu} (\partial_{\nu}h T_{h} - 
ig Z_{\nu}h [T_{Z}, T_{h}]) \noindent \\ 
&=& (\partial_{\mu}\partial_{\nu}h) T_{h} - ig \{Z_{\mu}\partial_{\nu}h
+ Z_{\nu}\partial_{\mu}h + (\partial_{\mu}Z_{\nu})h\}[T_{Z}, T_{h}] 
-g^{2} Z_{\mu}Z_{\nu}h[T_{Z}, [T_{Z}, T_{h}]].  
\label{0.9}   
\eea 
Let us note that the antisymmetric part under $\mu \leftrightarrow \nu$ 
is identical to (\ref{0.8}), as it should be from the Bianchi identity 
$D_{y}F_{\mu \nu} + D_{\mu}F_{\nu y} + D_{\nu}F_{y \mu} = 0$. 
Since (\ref{0.9}) does not contain the photonic field, 
${\rm Tr}\{ (D_{\mu}F_{\nu y})(D^{\mu}F^{\nu y})\}$ does not to
contribute to $H \to \gamma \gamma$ nor $H \to Z \gamma$. 

\noindent (3) ${\rm Tr}\{ (D_{y}F_{\mu y})(D^{y}F^{\mu y})\}$  
 
\be 
D_{y}F_{\mu y} = -igh [T_{h}, \partial_{\mu}h T_{h} - ig Z_{\mu}h [T_{Z}, T_{h}]]  
= -g^{2}Z_{\mu}h^{2}[T_{h}, [T_{Z}, T_{h}]].  
\label{0.9'}   
\ee 
Again, ${\rm Tr}\{ (D_{y}F_{\mu y})(D^{y}F^{\mu y})\}$ does
not contribute to $H \to \gamma \gamma$ nor to $H \to Z \gamma$. 

We thus have shown that there is no $d = 6$ gauge-invariant 
operator responsible for $H \to \gamma \gamma$ or $H \to Z \gamma$. 
The argument above can be generalized 
to gauge-invariant operators with arbitrary mass dimension.
The building block of the gauge-invariant operators is the successive
operation of the covariant derivative to the field strength, 
$D_{L_{1}}\cdots D_{L_{n}} F_{MN}$, the generalization of 
$D_{L}F_{MN}$. 
Suitably taking Tr of the product of these building blocks we get 
various gauge-invariant operators. 

Let us now investigate what kinds of operators we obtain from
$D_{L_{1}}\cdots D_{L_{n}} F_{MN}$. (As a specific case we 
include the situation where there is no covariant derivative.) 
We retain only the terms up to linear in $Z_{\mu}$, since we are
interested in the decays with no or one $Z$ boson in the final state. 
We also note that $D_{y} =-igh [T_{h}, $ when operating to the KK zero modes and 
$D_{\mu} = \partial_{\mu} - ig Z_{\mu} [T_{Z}, $ when operating to neutral fields . 

We classify into two cases depending on the choice of 
the indices $M, \ N$. 

\noindent (1) $D_{L_{1}}\cdots D_{L_{n}} F_{\mu \nu}$ 

We first note $F_{\mu \nu} = (\partial_{\mu} \gamma_{\nu} - 
\partial_{\nu}\gamma_{\mu})T_{\gamma} + (\partial_{\mu} Z_{\nu} - 
\partial_{\nu}Z_{\mu})T_{Z}$. 
So we consider each of the following two cases: 

\noindent (1a) $D_{L_{1}}\cdots D_{L_{n}} (\partial_{\mu} \gamma_{\nu} - 
\partial_{\nu}\gamma_{\mu})T_{\gamma}$ 

As the operation of $D_{y}$ to the photonic field $\gamma_{\mu}$ or its 
space-time derivatives yields a vanishing result because of $[T_{h}, T_{A}] = 0$, the
only possible type of operator in this case is 
\be 
\label{0.10}
D_{\mu_{1}}\cdots D_{\mu_{n}} (\partial_{\mu} \gamma_{\nu} - 
\partial_{\nu}\gamma_{\mu}) T_{\gamma} = \partial_{\mu_{1}} 
\cdots \partial_{\mu_{n}}(\partial_{\mu} \gamma_{\nu} - \partial_{\nu}\gamma_{\mu})T_{\gamma}. 
\ee 

\noindent (1b) $D_{L_{1}}\cdots D_{L_{n}} (\partial_{\mu} Z_{\nu} - 
\partial_{\nu}Z_{\mu})T_{Z}$ 

In this case, again the operation of $D_{\mu_{i}}$ is equivalent to
$\partial_{\mu_{i}}$, since we are interested in the operators up 
to linear in $Z_{\mu}$ and $[T_{\gamma}, T_{Z}] = [T_{\gamma}, T_{h}] = 0$. 
Also noting that $D_{y}$ is equivalent to `$-igh [T_{h},$'\,, we get operators of the form  
\be 
\label{0.11} 
h^{m}\partial_{\mu_{1}}\cdots \partial_{\mu_{l}} (\partial_{\mu} Z_{\nu} 
- \partial_{\nu}Z_{\mu}) 
[T_{h}, [T_{h}, \cdots [T_{h}, T_{Z}]\cdots ] ].
\ee  

\noindent (2) $D_{L_{1}}\cdots D_{L_{n}} F_{\mu y}$ 

We note $F_{\mu y} = \partial_{\mu}h T_{h} - ig Z_{\mu}h [T_{Z}, T_{h}]$
as in (\ref{0.9}). 
So we consider each of the following two cases: 

\noindent (2a) $D_{L_{1}}\cdots D_{L_{n}} (\partial_{\mu}h) T_{h}$ 

If there appears $D_{y}$ in $D_{L_{1}}\cdots D_{L_{n}}$, 
there should be one and only one $D_{\mu}$ behaving as 
$-igZ_{\mu}[T_{Z}, $ to the right of the rightmost $D_{y}$.
Otherwise, the operation of $D_{y}$ to $(\partial_{\mu}h) T_{h}$ 
yields $[T_{h}, T_{h}] = 0$. Thus what we obtain in this case
is either operators containing only the Higgs field  
\be 
\label{0.12}
\partial_{\mu_1} \cdots \partial_{\mu_n}\partial_{\mu}h T_{h}
\ee
or operators of the form  
\be 
\label{0.13}
h^{m}Z_{\mu}[T_{h}, [T_{h}, \cdots [T_{Z}, T_{h}]\cdots ] ], 
\ee 
where possible space-time derivatives acting on the fields have been suppressed for simplicity. 

\noindent (2b) $D_{L_{1}}\cdots D_{L_{n}} (Z_{\mu}h) [T_{Z}, T_{h}]$ 
 
The covariant derivative $D_{\mu}$ among $D_{L_{1}}\cdots D_{L_{n}}$
is equivalent to $\partial_{\mu}$, as we are not interested in 
the operators containing $Z_{\mu}$ more than one time, while $D_y$ 
behaves as `$-ig h[T_{h},$'. We thus get 
operators of the same form 
as the one in (\ref{0.13}). 

A few remarks are now in order. First, we have checked that the results above on the possible types of operators also can be confirmed by use of the method of mathematical induction. As the second remark, it may be interesting to note that the 
generators $T_{h} = \frac{1}{2}\lambda_{6}$ and 
$T_{Z} = \frac{1}{4}\lambda_{3} - \frac{\sqrt{3}}{4}\lambda_{8}$ 
act only on the lower two components of the fundamental triplet
representation of SU(3), i.e., as if they are the generators of 
the subgroup SU(2) acting as 
Pauli matrices $\sigma_{1}$ and $\sigma_{3}$, respectively. 
In this viewpoint, $T_{\gamma} = \frac{\sqrt{3}}{4}\lambda_{3} +
\frac{1}{4}\lambda_{8} = \frac{\sqrt{3}}{2}Q$ behaves as unit 
matrix $I_{2}$ in the subspace: 
\be 
\label{0.14}
T_{\gamma} \sim I_{2}, \ \ T_{h} \sim \sigma_{1}, \ \ T_{Z} \sim \sigma_{3}, 
\ee
where multiplied constant factors have been suppressed. This leads to 
\be 
\label{0.15}
[T_{h}, [T_{h}, \cdots [T_{h}, T_{Z}]\cdots ] ] \sim \sigma_{2} 
\ {\rm or} \ \sigma_{3}, 
\ee
depending on whether the number of the action of $T_{h}$ is odd or even. 

Now it is easy to see whether relevant operators exist for 
$H \to \gamma \gamma$ and $H \to Z \gamma$. 
Concerning the photonic decay $H \to \gamma \gamma$, we 
should take two operators of the type (\ref{0.10}) and 
one operator of the type (\ref{0.12}). 
Let us note that we cannot take the operator of the type (\ref{0.12}) 
more than once, since $\partial_{\mu}h$ is proportional to the physical Higgs field 
$H$ even after the replacement, $h \to v + H$. Multiplying these operators and then 
taking Tr, and by use of 
the fact, ${\rm Tr}(T_{\gamma}^{2}T_{h}) \sim {\rm Tr} \sigma_{1} = 0$, we conclude that 
no gauge-invariant local operator exists to describe $H \to \gamma \gamma$. 
We may also argue, relying on the gauge symmetry of the SM, especially SU(2)$_{L}$ symmetry, 
that an operator with two photonic fields and one Higgs field, as is seen in (\ref{0.1}), 
is not allowed. We will argue later that when the effects of global operators are taken 
into account, we get the $d = 6$ operator responsible for 
the photonic decay.  

Concerning another decay $H \to Z \gamma$, we should take one operator of the 
type (\ref{0.10})
, one operator of the type (\ref{0.11}) or (\ref{0.13}), and at most one operator of
the type (\ref{0.12}), and we should multiply all of them, finally taking Tr.  
We now realize that after taking Tr the operator vanishes, 
since [\,from (\ref{0.14}) and (\ref{0.15})\,]  
\bea  
&&{\rm Tr}(T_{\gamma} [T_{h}, [T_{h}, \cdots [T_{h}, T_{Z}]\cdots ]])
\sim {\rm Tr}(I_{2}\sigma_{3}) = 0, \nonumber \\  
&&{\rm Tr}(T_{\gamma} T_{h} [T_{h}, [T_{h}, \cdots [T_{h}, T_{Z}]\cdots ]]) 
\sim {\rm Tr}(I_{2}\sigma_{1}\sigma_{2}) = 0, 
\label{0.17} 
\eea 
where we have used the fact that we should have even numbers of $T_{h}$, 
in order to guarantee the SU(2)$_L$ gauge invariance. 
Thus we finally conclude that 
no gauge-invariant local operator exists
to describe $H \to Z \gamma$.

\subsection{The contributions of global operators} 

As a characteristic feature of the GHU models compactified on 
non-simply-connected extra space such as a circle, the Wilson loop 
$W = {\rm exp}(ig \oint A_{y}dy)$ due to the KK zero mode of $A_{y}$, 
namely, the Higgs field, makes physical sense. We now expand our argument 
to include this gauge-covariant global operator in the operator analysis. 
As a matter of fact, in 5D GHU the Higgs potential is induced at the 
quantum level and is written in terms of $W$. 

Including the Wilson loop, the simplest possibility to get the gauge-invariant
operator responsible for the Higgs decays of our interest should be 
\be 
\label{0.18}
{\rm Tr} (W^{m}) {\rm Tr}(F_{MN}F^{MN}) \ \ \to \ \
{\rm Tr}(A_{y}^{2}) {\rm Tr}(F_{\mu \nu}F^{\mu \nu}) \ \ \to \ \ 
h^{2} \{2{\rm Tr}(W_{\mu \nu}W^{\mu \nu}) + B_{\mu \nu}B^{\mu \nu}\}.  
\ee 
Thus obtained, the $d = 6$ operator clearly contributes to the photonic decay
but not to $H \to Z \gamma$, since 
\be 
\label{0.19}  
2{\rm Tr}(W_{\mu \nu}W^{\mu \nu}) + B_{\mu \nu}B^{\mu \nu} 
\ \ \to \ \ F^{(\gamma)}_{\mu \nu}F^{(\gamma) \mu \nu} + 
(\partial_{\mu} Z_{\nu} - \partial_{\nu}Z_{\mu})(\partial^{\mu}Z^{\nu}
- \partial^{\nu}Z^{\mu}), 
\ee 
when it is rewritten in terms of $\gamma_{\mu}$ and $Z_{\mu}$. 
More generally, we can think of operators of the form 
${\rm Tr} (W^{m}) \times$ (gauge-invariant local operators) with arbitrary mass dimension. 
By similar argument as the one used for the purely local operators,
we again exclude the possibility to get an operator responsible for $H \to Z \gamma$. 

To summarize the conclusion in this section, we have found the $d = 6$ 
operator for $H \to \gamma \gamma$ by taking the effect of the global 
operator into account, while the possibility of the gauge-invariant 
bulk operator responsible for $H \to Z \gamma$ is completely excluded. 

The absence of the relevant operator explains why the contribution 
of nonzero KK modes to $H \to Z\gamma$ exactly vanishes in the minimal SU(3) GHU 
model \cite{Maru:2013bja}. 
Let us make a brief comment on how the argument extended above may change in the
case of the SO(5)$\times$U(1) GHU model. 
In this model, since the gauge group is not a simple group, the Wilson coefficients 
of $2h^{2}{\rm Tr}(W_{\mu \nu}W^{\mu \nu})$ and $h^{2}B_{\mu \nu}B^{\mu \nu}$ in 
(\ref{0.18}) are anticipated to be different in general, which in turn means that 
when an orthogonal transformation to the base of $\gamma_{\mu}$ and $Z_{\mu}$ is made, 
the operators yield an operator responsible for $H \to Z \gamma$, i.e., 
$h^{2} F^{(\gamma)}_{\mu \nu}Z^{\mu \nu}$. This may be a reason why the contribution of 
the nonzero KK modes to the decay is not forbidden in the SO(5)$\times$U(1) 
GHU model \cite{Funatsu:2015xba}.

It may be interesting to ask whether the absence of the relevant operator for 
$H \to Z\gamma$ in the minimal SU(3) model also means that the contribution of
the KK zero mode also vanishes or not. It is interesting to note that in the SU(3) model 
the contribution of the KK zero mode of the SU(3) triplet fermion 
$(u_{L}, d_{L}, d_{R})$ (for the first generation) to $H \to Z\gamma$ is known to vanish, 
since the decay amplitude is proportional 
to $T_{3} - 2e_{f} \sin^{2}\theta_{W}$, where $T_{3}$ denotes the weak
isospin and the $e_{f}$ is the charge of the fermion \cite{Gunion:1989we}.
In this simplified model $u_{L}$ does not couple with the Higgs
field and 
$\sin^{2}\theta_{W} = \frac{3}{4}$, while $T_{3} = -\frac{1}{2}, \ e_{f} = - \frac{1}{3}$ 
for the contribution of the $d$ quark. 

We, however, naively expect that the contribution of the KK zero mode 
just recovers that of the SM. In fact, the contribution of the KK zero mode of 
$A_{\mu}$, i.e., 
the $W^{\pm}$ boson, to $H \to Z \gamma$ is known to be nonvanishing, even for 
$\sin^{2} \theta_{W} = \frac{3}{4}$ \cite{Gunion:1989we}. So it should be reasonable
to expect 
that in a realistic model with the realistic weak mixing angle, the SM
prediction is recovered. 

We also note that the orbifolding breaks the SU(3) gauge symmetry in the bulk into 
the symmetry of the 
SM in the sector of the KK zero mode of $A_{\mu}$. This may be the reason why the $W^{\pm}$ 
boson contributes to the decay, since the KK zero mode of $A_{\mu}$ contains only the 
gauge bosons of the SM [\,the ``incomplete multiplet" of SU(3)\,]. 
Hence, the contribution is 
not restricted by the operator analysis done above relying on the SU(3) symmetry.  
On the other hand, the quarks form a ``complete multiplet" of SU(3) triplet, if we 
ignore the difference of their chiralities, and their contributions respect the SU(3) 
symmetry. In the language of operators, the orbifolding may cause brane-localized operators, 
which respect only the gauge symmetry of the SM, as the contribution of the KK zero mode. 
As a matter of fact, the operators including the Wilson loop 
${\rm Tr} (W^{m}) {\rm Tr}(W_{\mu \nu}W^{\mu \nu}) \to h^{2}{\rm Tr}(W_{\mu \nu}W^{\mu \nu})$
and ${\rm Tr} (W^{m}) B_{\mu \nu}B^{\mu \nu} \to h^{2}B_{\mu \nu}B^{\mu \nu}$ may be 
generated with different Wilson coefficients, in general, which result in the operator 
responsible for the $H \to Z \gamma$ decay, after the orthogonal transformation into 
the base of $\gamma_{\mu}$ and $Z_{\mu}$.

\subsection{On the matter content of the model} 

Let us note that the results on the Higgs decay obtained above, relying on the 
gauge-invariant operators, do not refer to the matter content of the SU(3) model and are 
expected to hold irrespectively of the detail of the content. However, for the model 
to be viable, it is of crucial importance whether the model can incorporate quarks 
and leptons as the matter fields. 

As has already been mentioned above, each component of the SU(3) triplet has 
a fractional electric charge and the triplet fermions are identified with a quark 
multiplet: $(u_{L}, d_{L}, d_{R})$ (for the first generation). Then an important question 
is whether we can incorporate leptons with integer charges. Since the fundamental 
representation has fractional charges, $\frac{2}{3}, \ -\frac{1}{3}$, it 
will be natural to expect that the third-rank totally symmetric tensor representation, 
i.e., 10-dimensional tensor representation, has integer charges. In fact, it is easy to 
know that the representation contains SU(2)$_L$ doublet and singlet, which can be identified 
with $(\nu_{e L}, e^{-}_{L})$ and $e^{-}_{R}$, respectively
(again, for the first generation). The SU(2)$_L$ doublet also may be incorporated
in the adjoint 8 representation, though whose SU(2)$_L$ singlet component does not carry 
electric charge and cannot be identified with $e^{-}_{R}$. In any case, 
in addition to the desired leptonic fields, some exotic fields appear and 
a mechanism to remove such exotic states will be needed. A possible mechanism is 
to introduce brane-localized fermions, which form massive Dirac fields together 
with the exotic fields and decouple from the low-energy effective theory \cite{Burdman}.   

Another way to adjust the electric charges of the leptonic sector is, instead 
of introducing a higher-dimensional representation of SU(3), to add a U(1) factor, 
thus making 
the gauge group, e.g., SU(3) $\times$ U(1). We, however, would like to point out 
that in this case, just as in the case of SO(5) $\times$ U(1), the operator 
responsible for $H \to Z \gamma$ appears and the decay is not strictly forbidden.

\section{The contribution of nonzero KK modes of light quarks to the $H \to \gamma \gamma$ 
decay}

In this section we discuss the possible important contributions of 
nonzero KK modes of light quarks to the photonic decay $H \to \gamma \gamma$. 

We first give a generic formula for the contribution of a fermion with
the mass $m$, the charge $eQ$, and Yukawa coupling $f$. The contribution 
of the fermion to the decay amplitude can be written in the form of the 
effective Lagrangian, whose operator has appeared in (\ref{0.1}): 
\be 
\label{1.6} 
{\cal L} = C_{f} \ H F^{(\gamma)}_{\mu \nu}F^{(\gamma)\mu \nu}, 
\ee 
where the Wilson coefficient is given as \cite{Gunion:1989we}
\be  
C_{f} = -\frac{f\alpha Q^{2}}{2\pi m} \cdot I(\tau),  
\label{1.8} 
\ee  
with $\alpha = e^{2}/(4\pi)$. The function is defined as    
\be 
\label{1.3} 
I(\tau) = -\frac{\tau}{2} + \frac{\tau(\tau-1)}{2}(\sin^{-1}
\frac{1}{\sqrt{\tau}})^{2},  
\ee 
with 
\be 
\label{1.4}
\tau = \frac{4m^{2}}{m_{h}^{2}}. 
\ee 
In the limit $\tau \to \infty$, i.e., $m \gg m_{h}$, the function is well approximated by  
\be 
\label{1.5} 
I(\tau) \simeq  -\frac{\tau}{2} + \frac{\tau(\tau-1)}{2}(\frac{1}{\sqrt{\tau}}
+ \frac{1}{6}\frac{1}{(\sqrt{\tau})^{3}})^{2} \ \to \ 
I(\infty) = - \frac{1}{3}.  
\ee

Now we are ready to calculate the contribution of nonzero KK modes of
a light quark with charge $Q$ in the GHU scenario. 
The Yukawa coupling of its KK zero mode is exponentially 
suppressed by a factor $e^{-\pi MR} \ (M\ \mbox{is the bulk mass and} \ R\
\mbox{is the radius of} \ S^{1})$ compared with $g_{4} \ (\mbox{the 4D gauge coupling})$ 
to realize the light quark. 
Such exponential suppression is due to the localization of mode functions of 
left- and right-handed fermions at the different fixed points of the orbifold depending 
on their chiralities, caused by 
the presence of the ``$Z_{2}$-odd" bulk mass term.  

The motivation of our study concerning the contribution  
of nonzero KK modes is the fact that their Yukawa couplings are no longer exponentially 
suppressed as in the case of the KK zero mode, since their  
mode functions are not localized at the fixed points, behaving as 
ordinary trigonometric functions, roughly speaking. 
Thus their Yukawa couplings are comparable to the gauge coupling. Therefore, there is the 
chance to get a contribution to the Wilson coefficient
$C_{f}$ of Eq.(\ref{1.8}) from the nonzero KK modes of light quarks, which is comparable 
to that of the nonzero KK modes of the t quark. 

We will reasonably assume that the masses of nonzero KK fermions 
denoted by $m^{(\pm)}_{n} \ (n = 1, 2, \ldots)$ are much greater than the Higgs mass, 
\be 
\label{1.9} 
m_{h} \ll m^{(\pm)}_{n}.   
\ee 
As was discussed in \cite{Hasegawa:2012sy,Maru:2007xn,Lim:2007ae},
in GHU scenario the $n$th
mass eigenvalue splits into two eigenvalues by the effect of the Higgs VEV: 
\be 
\label{1.10} 
m^{(\pm)}_{n} = \frac{n}{R} \pm M_{W} 
\ee 
with $M_{W} = \frac{g_{4}}{2}v$ for the simplified case of the vanishing bulk mass, $M = 0$. 

Under (\ref{1.9}), the coefficient $C^{(\pm)}_{n}$ denoting the contribution
of the KK mode with the mass $m^{(\pm)}_{n}$ can be written
[\,by use of (\ref{1.5}) and (\ref{1.8})\,] as  
\be 
\label{1.11} 
C^{(\pm)}_{n} \simeq \frac{f_{n}^{(\pm)}\alpha Q^{2}}{6\pi m^{(\pm)}_{n}},   
\ee 
where $f_{n}^{(\pm)}$ is the Yukawa coupling of the KK 
mode, which is generally obtained by taking the derivative 
of the mass eigenvalue with respect to the VEV \cite{Hasegawa:2012sy}, 
\be 
\label{1.12} 
f_{n}^{(\pm)} = \frac{\partial m^{(\pm)}_{n}}{\partial v} = 
\frac{g_{4}}{2} \frac{\partial m^{(\pm)}_{n}}{\partial M_{W}}. 
\ee 

Thus, what we have to do is to perform the KK mode sum 
\be 
\label{1.13} 
\sum_{n = 1}^{\infty} \ \frac{f_{n}^{(\pm)}}{m^{(\pm)}_{n}} ( =
\sum_{n=1}^{\infty} \ \frac{g_{4}}{2} \frac{\partial \log \
m^{(\pm)}_{n}}{\partial M_{W}}). 
\ee 
In the SU(3) GHU model with the bulk mass $M$, as was discusses in \cite{Hasegawa:2012sy},
the mass eigenvalue is the solution of 
\be 
\label{1.14} 
\pm (-1)^{n} \sin (M_{W} \pi R) = \frac{m^{(\pm)}_{n}}{\sqrt{m^{(\pm)2}_{n}
- M^{2}}} \ \sin (\sqrt{m^{(\pm)2}_{n} - M^{2}} \pi R).  
\ee 
Although this equation cannot be solved analytically for
$m^{(\pm)}_{n}$, it is still possible to get the eigenvalue approximately by 
utilizing the perturbative 
expansion in terms of $M_{W}$. 
Ignoring the ${\cal O}(M_{W}^{3})$, $m^{(\pm)}_{n}$ is written as   
\be 
\label{1.15} 
m^{(\pm)}_{n} \simeq \sqrt{(\frac{n}{R})^{2} + M^{2}} + 
\alpha_{n}^{(\pm)} M_{W} + \beta_{n}^{(\pm)} M_{W}^{2}. 
\ee 
The reason to keep the terms up to
${\cal O}(M_{W}^{2})$ is that the operators in (\ref{0.2}) suggest that the contribution to the decay amplitude appears only at the second order of the weak scale. 

From (\ref{1.15}), we get the following approximate relations:
\bea  
\sin (\sqrt{m^{(\pm)2}_{n} - M^{2}} \pi R) &\simeq& (-1)^{n}
\{ \frac{\sqrt{(\frac{n}{R})^{2} + M^{2}}}{(\frac{n}{R})}\alpha_{n}^{(\pm)}
(M_{W}\pi R) 
+ \frac{\sqrt{(\frac{n}{R})^{2} + M^{2}}}{n \pi} \beta_{n}^{(\pm)}
(M_{W}\pi R)^{2} \nonumber \\ 
&& - \frac{1}{2}\frac{M^{2}}{n\pi (\frac{n}{R})^{2}} \alpha_{n}^{(\pm)2} 
(M_{W}\pi R)^{2} \}. 
\label{1.16} 
\eea 
Substituting these approximate relations in the rhs of (\ref{1.14}),
we get 
\bea 
&& \frac{m^{(\pm)}_{n}}{\sqrt{m^{(\pm)2}_{n} - M^{2}}} \ 
\sin (\sqrt{m^{(\pm)2}_{n} - M^{2}} \pi R) 
\simeq (-1)^{n} \frac{1}{(\frac{n}{R})}\times 
\{ \frac{(\frac{n}{R})^{2} + M^{2}}{(\frac{n}{R})} \alpha_{n}^{(\pm)} 
(M_{W}\pi R) \nonumber \\ 
&& + \frac{1}{n\pi}[-\frac{3}{2}\frac{M^{2}}{(\frac{n}{R})^{2}}
\sqrt{(\frac{n}{R})^{2}+ M^{2}} \alpha_{n}^{(\pm)2} + ((\frac{n}{R})^{2} + M^{2})
\beta_{n}^{(\pm)}]
(M_{W}\pi R)^{2} \}. 
\label{1.17}
\eea 
On the other hand, the lhs of (\ref{1.14}) is simply approximated 
up to the ${\cal O}(M_{W}^{2})$ as 
\be 
\label{1.18}  
\pm (-1)^{n} \sin (M_{W} \pi R) \simeq \pm (-1)^{n} (M_{W} \pi R).  
\ee 
Comparing this with (\ref{1.17}), we get coupled equations for the 
coefficients $\alpha_{n}^{(\pm)}$, $\beta_{n}^{(\pm)}$: 
\bea 
&& \frac{(\frac{n}{R})^{2} + M^{2}}{(\frac{n}{R})^{2}} \alpha_{n}^{(\pm)} = \pm 1,  
\label{1.19} \\  
&& -\frac{3}{2}\frac{M^{2}}{(\frac{n}{R})^{2}}\sqrt{(\frac{n}{R})^{2}+ M^{2}}
\alpha_{n}^{(\pm)2} + ((\frac{n}{R})^{2} + M^{2})\beta_{n}^{(\pm)} = 0, 
\label{1.20} 
\eea 
which lead to 
\bea 
&& \alpha_{n}^{(\pm)} = \pm \frac{(\frac{n}{R})^{2}}{(\frac{n}{R})^{2} + M^{2}}, 
\label{1.21} \\   
&& \beta_{n}^{(\pm)} = \frac{3}{2}\frac{(\frac{n}{R})^{2}M^{2}}{[(\frac{n}{R})^{2}
+ M^{2}]^{\frac{5}{2}}}. 
\label{1.22}  
\eea 

We thus have obtained the approximate formula for the mass eigenvalues, 
\be 
\label{1.23} 
m_{n}^{(\pm)} \simeq \sqrt{(\frac{n}{R})^{2} + M^{2}} \pm 
\frac{(\frac{n}{R})^{2}}{(\frac{n}{R})^{2} + M^{2}} M_{W} + 
\frac{3}{2}\frac{(\frac{n}{R})^{2}M^{2}}{[(\frac{n}{R})^{2} +
M^{2}]^{\frac{5}{2}}} M_{W}^{2}. 
\ee 

Then the ratio of the Yukawa coupling to the mass eigenvalue is
approximated to be 
\bea   
&&\frac{f_{n}^{(\pm)}}{m^{(\pm)}_{n}} = \frac{\frac{g_{4}}{2} 
\frac{\partial m^{(\pm)}_{n}}{\partial M_{W}}}{m^{(\pm)}_{n}} 
\simeq \frac{g_{4}}{2} 
\frac{\alpha_{n}^{\pm} + 2\beta_{n}^{(\pm)}M_{W}}{\sqrt{(\frac{n}{R})^{2}+M^{2}}
+ \alpha_{n}^{(\pm)}M_{W}} \nonumber \\ 
&& \simeq \frac{g_{4}}{2} \{ \pm \frac{(\frac{n}{R})^{2}}{[(\frac{n}{R})^{2} 
+ M^{2}]^{\frac{3}{2}}} + \frac{3 (\frac{n}{R})^{2}M^{2} -
(\frac{n}{R})^{4}}{[(\frac{n}{R})^{2} + M^{2}]^{3}} M_{W} \}.  
\label{1.24} 
\eea 

The first term, corresponding to the possible $d =5$ operator, cancels out between the contributions of two different types of nonzero KK modes with $m_{n}^{(+)}$ and $m_{n}^{(-)}$, as it should be. Thus we are left with 
the summation 
\be 
\label{1.25} 
\sum_{n = 1}^{\infty} \ \frac{3 (\frac{n}{R})^{2}M^{2} -
(\frac{n}{R})^{4}}{[(\frac{n}{R})^{2} + M^{2}]^{3}},  
\ee 
corresponding to the contribution of the $d = 6$ operator, which should be UV-finite for 5D space-time.   
The sum may be rearranged into three terms: 
\be 
\label{1.26} 
\sum_{n = 1}^{\infty} \ \frac{3 (\frac{n}{R})^{2}M^{2} - 
(\frac{n}{R})^{4}}{[(\frac{n}{R})^{2} + M^{2}]^{3}} 
= - \sum_{n = 1}^{\infty} \frac{1}{(\frac{n}{R})^{2} + M^{2}} + 
5M^{2} \sum_{n = 1}^{\infty} \frac{1}{[(\frac{n}{R})^{2} + M^{2}]^{2}} 
- 4M^{4} \sum_{n = 1}^{\infty} \frac{1}{[(\frac{n}{R})^{2} + M^{2}]^{3}}.   
\ee 

The relevant formulas are the following, where the second and the third formulas are 
obtained by taking a derivative 
$- \frac{d}{da^{2}} = - \frac{1}{2a}\frac{d}{da}$ of the previous formulas: 
\bea 
&& \sum_{n = 1}^{\infty} \frac{1}{n^{2} + a^{2}} 
= -\frac{1}{2a^{2}} + \frac{\pi}{2a} \coth (a\pi),  
\label{1.27} \\ 
&& \sum_{n = 1}^{\infty} \frac{1}{(n^{2} + a^{2})^{2}} 
= -\frac{1}{2a^{4}} + \frac{\pi}{4a^{3}} \coth (a\pi) + 
\frac{\pi^{2}}{4a^{2}} \frac{1}{\sinh^{2} (a\pi)},  
\label{1.28} \\ 
&& \sum_{n = 1}^{\infty} \frac{1}{(n^{2} + a^{2})^{3}} 
= -\frac{1}{2a^{6}} + \frac{3}{16}\frac{\pi^{2}}{a^{4}} 
\frac{1}{\sinh^{2} (a\pi)} 
+ \frac{3}{16}\frac{\pi}{a^{5}} \coth (a\pi) + \frac{1}{8}
\frac{\pi^{3}}{a^{3}} \frac{\coth (a \pi)}{\sinh^{2} (a\pi)}. 
\label{1.29} 
\eea 
By use of these formulas, the KK mode summation is performed as, with $MR = a$,  
\bea 
&& \sum_{n = 1}^{\infty} \ \frac{3 (\frac{n}{R})^{2}M^{2} -
(\frac{n}{R})^{4}}{[(\frac{n}{R})^{2} + M^{2}]^{3}} \nonumber \\ 
&& = -R^{2}[-\frac{1}{2a^{2}} + \frac{\pi}{2a} \coth (a\pi)] \nonumber \\ 
&& + 5R^{2}a^{2} [-\frac{1}{2a^{4}} + \frac{\pi}{4a^{3}} \coth (a\pi) +
\frac{\pi^{2}}{4a^{2}} \frac{1}{\sinh^{2} (a\pi)}] \nonumber \\ 
&& - 4R^{2}a^{4} [-\frac{1}{2a^{6}} + \frac{3}{16}\frac{\pi^{2}}{a^{4}}
\frac{1}{\sinh^{2} (a\pi)} 
+ \frac{3}{16}\frac{\pi}{a^{5}} \coth (a\pi) + \frac{1}{8} \frac{\pi^{3}}{a^{3}}
\frac{\coth (a \pi)}{\sinh^{2} (a\pi)}] \nonumber \\ 
&& = \frac{\pi^{2}}{2}R^{2} \frac{1 - (a\pi)\coth (a\pi)}{\sinh^{2}(a\pi)}. 
\label{1.30} 
\eea 

Surprisingly, in (\ref{1.30}) the terms with power suppression,
i.e., the terms proportional to $\frac{R^{2}}{a^{2}}$ and $\frac{R^{2}\pi}{a}\coth (a\pi)$,
turn out to disappear completely. This means that the KK mode
sum leaves a contribution, which is exponentially suppressed for the case 
$\pi MR = a\pi \gg 1$: 
\be 
\label{1.32} 
\sum_{n = 1}^{\infty} \ \frac{3 (\frac{n}{R})^{2}M^{2} -
(\frac{n}{R})^{4}}{[(\frac{n}{R})^{2} + M^{2}]^{3}} \sim 2\pi^{2}R^{2}
(1 - \pi MR) e^{-2\pi MR}.  
\ee 
Thus, it has turned out that the contribution of the nonzero KK modes of 
light quarks to the photonic Higgs decay is strongly suppressed
compared to the contribution of the nonzero KK modes of the t quark by the factor 
$e^{-2\pi MR}$, in contrast to our naive expectation.  

Such exponential suppression may be understood from the operator analysis in 
the previous section. There we found that the $d = 6$ 
operator including the effect of the Wilson-loop, Eq.(\ref{0.18}), contributes
to the photonic decay. Now the exponential suppression, seen in (\ref{1.32}), is an 
inevitable consequence of the operator, since the Wilson loop comes from the Feynman diagrams 
where the fermion loop is wrapped around $S^{1}$ of the extra space, where 
the fermion propagator gets a suppression factor $e^{-MR}$, as in the case 
of the Yukawa-type potential, 
due to the presence of the bulk mass $M$. Or, in the analogous situation of 
finite temperature field theory, the factor $e^{-MR}$ can be understood to 
correspond to the Boltzmann factor $e^{-\frac{M}{T}}$. A similar exponential 
suppression factor was found in the effective Higgs potential in 
GHU \cite{Hatanaka:1998yp}, which 
is also described by the Wilson loop.

\subsection*{Acknowledgments}

We would like to thank N. Maru for informative and stimulating discussions on 
their pioneering works on the Higgs decays. Thanks are also due to N. Kurahashi 
for his enjoyable collaboration in the early stage of this work. 
The work of C.S.L. was supported in part by the Japan Society for the Promotion of 
Science, Grants-in-Aid for Scientific Research, No.~23104009 and No.~15K05062.

%\bibliography{./GHU_ref_library}

\begin{thebibliography}{10}

\bibitem{Aad:2012tfa}
{\bf ATLAS} Collaboration, G.~Aad et~al., {\it {Observation of a new particle
  in the search for the Standard Model Higgs boson with the ATLAS detector at
  the LHC}},  {\em Phys. Lett.} {\bf B716} (2012) 1--29,
  [\href{http://xxx.lanl.gov/abs/1207.7214}{{\tt arXiv:1207.7214}}].

\bibitem{Chatrchyan:2012xdj}
{\bf CMS} Collaboration, S.~Chatrchyan et~al., {\it {Observation of a new boson
  at a mass of 125 GeV with the CMS experiment at the LHC}},  {\em Phys. Lett.}
  {\bf B716} (2012) 30--61, [\href{http://xxx.lanl.gov/abs/1207.7235}{{\tt
  arXiv:1207.7235}}].

\bibitem{Manton:1979kb}
N.~S. Manton, {\it {A New Six-Dimensional Approach to the Weinberg-Salam
  Model}},  {\em Nucl. Phys.} {\bf B158} (1979) 141.

\bibitem{Hosotani:1983xw}
Y.~Hosotani, {\it {Dynamical Mass Generation by Compact Extra Dimensions}},
  {\em Phys. Lett.} {\bf B126} (1983) 309.

\bibitem{Hosotani:1983vn}
Y.~Hosotani, {\it {Dynamical Gauge Symmetry Breaking as the Casimir Effect}},
  {\em Phys. Lett.} {\bf B129} (1983) 193.

\bibitem{Hosotani:1988bm}
Y.~Hosotani, {\it {Dynamics of Nonintegrable Phases and Gauge Symmetry
  Breaking}},  {\em Annals Phys.} {\bf 190} (1989) 233.

\bibitem{Hatanaka:1998yp}
H.~Hatanaka, T.~Inami, and C.~S. Lim, {\it {The Gauge hierarchy problem and
  higher dimensional gauge theories}},  {\em Mod. Phys. Lett.} {\bf A13} (1998)
  2601--2612, [\href{http://xxx.lanl.gov/abs/hep-th/9805067}{{\tt
  hep-th/9805067}}].

\bibitem{Hosotani:2007qw}
Y.~Hosotani and Y.~Sakamura, {\it {Anomalous Higgs couplings in the 
SO(5)$\times$U$(1)_{B-L}$ gauge-Higgs unification in warped spacetime}},  {\em Prog. Theor.
  Phys.} {\bf 118} (2007) 935--968,
  [\href{http://xxx.lanl.gov/abs/hep-ph/0703212}{{\tt hep-ph/0703212}}].

\bibitem{Hosotani:2008tx}
Y.~Hosotani, K.~Oda, T.~Ohnuma, and Y.~Sakamura, {\it {Dynamical Electroweak
  Symmetry Breaking in SO(5)$\times$U(1) Gauge-Higgs Unification with Top and Bottom
  Quarks}},  {\em Phys. Rev.} {\bf D78} (2008) 096002,
  [\href{http://xxx.lanl.gov/abs/0806.0480}{{\tt arXiv:0806.0480}}]. [Erratum:
  Phys. Rev.D79,079902(2009)].

\bibitem{Hosotani:2008by}
Y.~Hosotani and Y.~Kobayashi, {\it {Yukawa Couplings and Effective Interactions
  in Gauge-Higgs Unification}},  {\em Phys. Lett.} {\bf B674} (2009) 192--196,
  [\href{http://xxx.lanl.gov/abs/0812.4782}{{\tt arXiv:0812.4782}}].

\bibitem{Hosotani:2009jk}
Y.~Hosotani, P.~Ko, and M.~Tanaka, {\it {Stable Higgs Bosons as Cold Dark
  Matter}},  {\em Phys. Lett.} {\bf B680} (2009) 179--183,
  [\href{http://xxx.lanl.gov/abs/0908.0212}{{\tt arXiv:0908.0212}}].

\bibitem{Funatsu:2013ni}
S.~Funatsu, H.~Hatanaka, Y.~Hosotani, Y.~Orikasa, and T.~Shimotani, {\it {Novel
  universality and Higgs decay H$\to \gamma\gamma$, gg in the SO(5)$\times$U(1)
  gauge-Higgs unification}},  {\em Phys. Lett.} {\bf B722} (2013) 94--99,
  [\href{http://xxx.lanl.gov/abs/1301.1744}{{\tt arXiv:1301.1744}}].

\bibitem{Funatsu:2014fda}
S.~Funatsu, H.~Hatanaka, Y.~Hosotani, Y.~Orikasa, and T.~Shimotani, {\it {LHC
  signals of the $SO(5)\times U(1)$ gauge-Higgs unification}},  {\em Phys.
  Rev.} {\bf D89} (2014), no.~9 095019,
  [\href{http://xxx.lanl.gov/abs/1404.2748}{{\tt arXiv:1404.2748}}].

\bibitem{Hasegawa:2012sy}
K.~Hasegawa, N.~Kurahashi, C.~S. Lim, and K.~Tanabe, {\it {Anomalous Higgs
  Interactions in Gauge-Higgs Unification}},  {\em Phys. Rev.} {\bf D87}
  (2013), no.~1 016011, [\href{http://xxx.lanl.gov/abs/1201.5001}{{\tt
  arXiv:1201.5001}}].

\bibitem{Maru:2007xn}
N.~Maru and N.~Okada, {\it {Gauge-Higgs unification at LHC}},  {\em Phys. Rev.}
  {\bf D77} (2008) 055010, [\href{http://xxx.lanl.gov/abs/0711.2589}{{\tt
  arXiv:0711.2589}}].

\bibitem{Maru:2008cu}
N.~Maru, {\it {Finite Gluon Fusion Amplitude in the Gauge-Higgs Unification}},
  {\em Mod. Phys. Lett.} {\bf A23} (2008) 2737--2750,
  [\href{http://xxx.lanl.gov/abs/0803.0380}{{\tt arXiv:0803.0380}}].

\bibitem{Kubo:2001zc}
M.~Kubo, C.~S. Lim, and H.~Yamashita, {\it {The Hosotani mechanism in bulk
  gauge theories with an orbifold extra space $S^{1}/Z_{2}$}},  {\em Mod. Phys.
  Lett.} {\bf A17} (2002) 2249--2264,
  [\href{http://xxx.lanl.gov/abs/hep-ph/0111327}{{\tt hep-ph/0111327}}].

\bibitem{Scrucca:2003ra}
C.~A. Scrucca, M.~Serone, and L.~Silvestrini, {\it {Electroweak symmetry
  breaking and fermion masses from extra dimensions}},  {\em Nucl. Phys.} {\bf
  B669} (2003) 128--158, [\href{http://xxx.lanl.gov/abs/hep-ph/0304220}{{\tt
  hep-ph/0304220}}].
  
\bibitem{Maru:2013bja}
N.~Maru and N.~Okada, {\it {$H \to Z\gamma$ in gauge-Higgs unification}},  {\em
  Phys. Rev.} {\bf D88} (2013), no.~3 037701,
  [\href{http://xxx.lanl.gov/abs/1307.0291}{{\tt arXiv:1307.0291}}].

\bibitem{Funatsu:2015xba}
S.~Funatsu, H.~Hatanaka, and Y.~Hosotani, {\it {H $\to$ Z$\gamma$ in the gauge-Higgs
  unification}},  {\em Phys. Rev.} {\bf D92} (2015), no.~11 115003,
  [\href{http://xxx.lanl.gov/abs/1510.0655}{{\tt arXiv:1510.0655}}].


\bibitem{Lim:2007ae}
C.~S. Lim and N.~Maru, {\it {Calculable One-Loop Contributions to S and T Parameters
                        in the Gauge-Higgs Unification}},  {\em Phys. Rev.} {\bf D75} (2007) 115011,
  [\href{http://xxx.lanl.gov/abs/0703.015}{{\tt arXiv:0703.015}}].

\bibitem{Gunion:1989we}
J.~F. Gunion, H.~E. Haber, G.~L. Kane, and S.~Dawson, {\it {The Higgs Hunter's
  Guide}},  {\em Front. Phys.} {\bf 80} (2000) 1--448.
  
\bibitem{Burdman}
G.~Burdman and Y.~Nomura, {\em Nucl. Phys.} {\bf B656} (2003) 3.

\end{thebibliography}
%\bibliographystyle{./JHEP}

%%%%%%%%%%%%%%%%%%%%
%%%% From here
%%%%%%%%%%%%%%%%%%%%
\providecommand{\href}[2]{#2}\begingroup\raggedright\endgroup
%%%%%%%%%%%%%%%%%%%%
%%%% Till here
%%%%%%%%%%%%%%%%%%%%

\end{document}